\documentclass[%
 reprint,
 amsmath,amssymb,
 aps,
 showpacs,
prl,
]{revtex4-1}

\usepackage{graphicx}
\usepackage{dcolumn}
\usepackage{bm}
\usepackage{hyperref}

\def\be{\begin{equation}}
\def\ee{\end{equation}}
\def\bea{\begin{eqnarray}}
\def\eea{\end{eqnarray}}
\def\bphi{\mbox{\boldmath $\phi$}}
\def\bsig{\mbox{\boldmath $\sigma$}}

\begin{document}


\title{Non-Fermi-liquid behavior and anomalous suppression of Landau damping in layered metals close to ferromagnetism} 

\author{Sam P. Ridgway}
\author{Chris A. Hooley}
\affiliation{%
Scottish Universities Physics Alliance, School of Physics and Astronomy, University of St Andrews,
North Haugh, St Andrews, Fife KY16 9SS, United Kingdom
}%

\date{\today}

\begin{abstract}
We analyse the low-energy physics of nearly ferromagnetic metals in two spatial dimensions using the functional renormalization group technique.  We find a new low-energy fixed point, at which the fermionic (electron-like) excitations are non-Fermi-liquid ($z_f = 13/10$) and the magnetic fluctuations exhibit an anomalous Landau damping whose rate vanishes as $\Gamma_{\bf q} \sim \vert {\bf q} \vert^{3/5}$ in the low-$\vert {\bf q} \vert$ limit.  We discuss this renormalization of the Landau-damping exponent, which is the major novel prediction of our work, and highlight the possible link between that renormalization and neutron-scattering data on UGe$_2$ and related compounds.  Implications of our analysis for YFe$_2$Al$_{10}$ are also discussed.
\end{abstract}

\maketitle


\textit{Introduction.} 
The problem of describing the low-temperature behavior of metals close to a magnetic instability is now several decades old, and many experimental examples are available.  These include the cuprate superconductors \cite{lee}, heavy fermion materials \cite{stew_rev,lohn_rev}, and nearly ferromagnetic metals \cite{pfleiderer}.  Among the phenomena observed in them are non-Fermi-liquid behavior of the conduction electrons, anomalous power laws in thermodynamic observables, and the emergence of new phases in the vicinity of the instability point.  That point is commonly modelled by coupling fermions with a gapless Fermi surface (the electrons) to a massless boson (representing the magnetic fluctuations).

The distinction between an instability to ferromagnetism and an instability to antiferromagnetism is an important one.  For incipient antiferromagnets, the magnetic fluctuations are peaked at some non-zero wavevector ${\bf Q}$, and scatter the electrons between certain `hot spots' on the Fermi surface.  In this Letter, however, we address the ferromagnetic case; furthermore, we work in two spatial dimensions, in which this is a strong-coupling problem.  Because the instability is ferromagnetic, the magnetic fluctuations are peaked at ${\bf Q}={\bf 0}$, which implies strong forward scattering at every point on the Fermi surface.  Nonetheless, the induced correlations between the fermions are strongest between points on the Fermi surface which have a common tangent.  These regions, known as `patches', are often (though not in this Letter) the only parts of the Fermi surface that are retained in theories of ferromagnetic quantum criticality \cite{polchinski,lee_gauge,met_i,mross,dalidovich,sur}.

Due to the quantum-mechanical effects that prevail at such low temperatures, one cannot separate the static and dynamic properties of the system.  To address this issue, Hertz put forward his theory of quantum criticality in the 1970s \cite{hertz}.  An important physical ingredient of the Hertz theory is Landau damping:\ the decay of magnetic fluctuations into quasiparticle-quasihole pairs. Hertz integrated out the fermions to produce a purely bosonic description of the quantum critical point \cite{hertz,millis}: the boson propagator is modified by the Landau damping term in the expansion of the particle-hole bubble, enforcing a bosonic dynamical exponent $z_b=3$, and the critical point is reduced to a conventional scalar field theory.  But the Hertz theory is unsatisfactory, as it assumes a Fermi-liquid form of the fermion propagator.  In addition, more careful analyses suggest that multi-paramagnon interactions become singular, rendering a purely bosonic description invalid \cite{hertz,thier}.

The failure of Hertz-type theories has motivated a concentration on theories that retain both the electronic (fermionic) and magnetic (bosonic) degrees of freedom in their low-energy description.  Such a theory may be treated by the standard Wilsonian renormalization approach \cite{fitz_wils,fitzpatricknew}; this results in a non-Fermi liquid fixed point in an expansion in $\epsilon = 3-d$.  However, this approach does not capture the onset of Landau damping, and thus gives physically incorrect results at low energies.

An alternative approach is to calculate the one-loop fermionic self energy using the Landau-damped propagator of Hertz theory.  The one-loop dynamic self-energy scales as $\omega^{2/3}$ \cite{holstein,reizer,polchinski,altshuler}, dominating the bare dynamic term $\omega$ in the low-energy limit.  Despite the fact that the procedure used to obtain this result is clearly not self-consistent, the $\omega^{2/3}$ form of the self-energy was argued to be exact by Rech {\it et al.\/}\ \cite{rech}, via an Eliashberg approach which is controlled by a large-$N_f$ limit.  However, Sung-Sik Lee \cite{lee_gauge} showed that a class of planar diagrams causes the large-$N_f$ methods to fail below a certain energy scale; this was confirmed by explicit three-loop calculations by Metlitski and Sachdev \cite{met_i}.  Thus, despite recent intensive work \cite{mross,dalidovich,sur,torroba}, the problem of fermions interacting with an overdamped critical bosonic mode remains unsolved.

In this work, we present a functional renormalization group (fRG) analysis of the ferromagnetic critical point.  We assume the existence of a continuous phase transition, ignoring questions of the stability of the fixed point \cite{chubukov2004,belitz2005,rech}.
Our treatment explicitly includes Landau damping, using a method recently applied by Lee, Strack, and Sachdev to an antiferromagnetic quantum critical point in a lattice model \cite{af_frg}.  It relies neither on a patch method nor on a large-$N_f$ expansion \cite{drukier}:\ we retain the full Fermi surface at all stages in the flow.  This produces vertex corrections that are absent in the patch approach, resulting in a novel dependence of the Landau damping rate on the wavevector of the magnetic fluctuations.  This surprising result is the chief novel prediction of our work.

\textit{Flow equations.} We begin with a model of spin-$1/2$ fermions $\psi^\alpha$ with a circular Fermi surface, and couple these to a bosonic field $\bphi$ via a Yukawa coupling $g$:
\begin{eqnarray}
S & = & - \int\limits_{\omega{\bf k}} \bar{\psi}^\alpha_{\omega {\bf k}} \left( i\omega - \tilde{k} \right) \psi^\alpha_{\omega {\bf k}} + \frac{1}{2} \int\limits_{\Omega {\bf q}} (|\textbf{q}|^2 + m^2) \bphi^{*}_{\Omega {\bf q}} \cdot \bphi_{\Omega {\bf q}} \nonumber \\ & &
\qquad + \,g \int\limits_{\omega {\bf k}} \int\limits_{\Omega {\bf q}} \bar{\psi}^\alpha_{\omega+\Omega,{\bf k}+{\bf q}} \bsig^{\alpha\beta} \psi^\beta_{\omega {\bf k}} \cdot \bphi_{\Omega {\bf q}} \,. \label{ba}
\end{eqnarray}
Here $\alpha$ and $\beta$ are spin indices, $\tilde{k} \equiv |\textbf{k}|-k_F$ describes the linearized dispersion of the fermions near the Fermi momentum $k_F$, and $m$ is the boson mass, which is set to zero since we work at criticality. We use the notation
$
\int_{\omega{\bf k}} \,\, = \,\,(2 \pi)^{-3} \int_{-\infty}^{\infty} \mathrm{d} \omega  \int^{|\tilde{k}| < \Lambda_{\rm UV}}  \mathrm{d}^{2} \textbf{k}
$
and
$
\int_{\Omega{\bf q}} \,\, = \,\, (2 \pi)^{-3} \int_{-\infty}^{\infty} \mathrm{d} \Omega  \int^{|\textbf{q}|<\Lambda_{\rm UV}} \mathrm{d}^{2} \textbf{q}$, where $\Lambda_{\mathrm{UV}} \ll k_F$ is an ultraviolet cutoff.
We do not include any dynamics of the boson in the bare action (\ref{ba}); the dynamics will be generated via the interactions with the fermions.

As usual in fRG, we follow the flow of the generating functional of one-particle irreducible vertex functions $\Gamma^\Lambda_R$ as the infrared scale $\Lambda$ is reduced from $\Lambda_{\rm UV}$ to zero \cite{wett,morris,rev_frg}:
\begin{equation}
\frac{\mathrm{d}}{\mathrm{d} \Lambda}\Gamma_{R}^{\Lambda}  = \frac{1}{2} \mathrm{STr} \left( \partial_{\Lambda}R^{\Lambda} \left[ \Gamma_{R}^{(2) \Lambda} + R^{\Lambda}\right]^{-1} \right)  \,.
\end{equation}
The generating functional $\Gamma^\Lambda_R$ flows from the microscopic action $\Gamma^{\Lambda_{\mathrm{UV}}}_R = S$ to the effective action $\Gamma^{\Lambda \rightarrow 0}_R = \Gamma$.  Our truncation of $\Gamma^\Lambda_R$ includes only the dressed fermionic  propagator $G^{\Lambda}_f $, the dressed bosonic propagator $D^{\Lambda}_b $, and the Yukawa coupling $g^{\Lambda}$. The regulator $R^{\Lambda}$ cuts off the infrared divergences in the fermionic and bosonic propagators; however, in order to capture the physics of Landau damping --- which is generated only by low-energy fermions --- we must follow Lee, Strack, and Sachdev \cite{af_frg} and set the fermionic regulator function to zero.

The fermionic matrix element of $[ \Gamma_{R}^{(2) \Lambda} + R^{\Lambda} ]^{-1} \big\vert_{\mbox{\footnotesize \boldmath $\phi$}={\bf 0}}$ is
\begin{equation}
G^{\Lambda}_f (\omega,{\bf k}) = \left({i\omega - \tilde{k} - \Sigma_f^{\Lambda}(\omega,{\bf k})} \right)^{-1} \,,
\end{equation}
where $\Sigma_f^{\Lambda}(\omega,{\bf k})$ represents the fermion self energy. We will only be interested in the low energy behavior of the model and so will parameterize the self energy as
\be
\Sigma_f^\Lambda(\omega,{\bf k}) = \left( 1 - A_{\omega}^\Lambda \right) i\omega - \left( 1 - A_k^\Lambda \right) {\tilde k}, \label{fermiself}
\ee
where the Fermi momentum has been kept fixed. We may use these parameters to calculate the renormalized Fermi velocity, $v_f^\Lambda = A_k^\Lambda/A_{\omega}^\Lambda$, and the quasiparticle weight, $Z^{\Lambda}_f=1/{A}^{\Lambda}_\omega$, at all stages of the flow. 

\begin{figure}[t]
(a) \includegraphics[
  width=3.2cm]{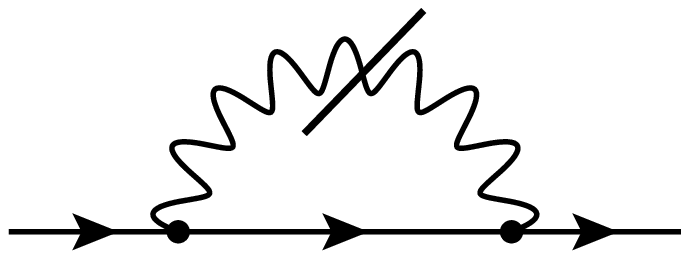} \\ \vspace*{5mm} (b) \includegraphics[
  width=3.2cm]{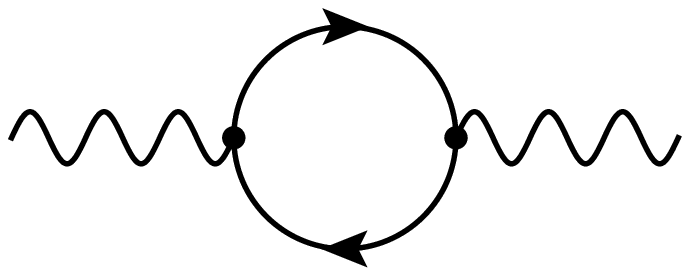} \hspace*{1cm} (c) \includegraphics[
  width=3.2cm]{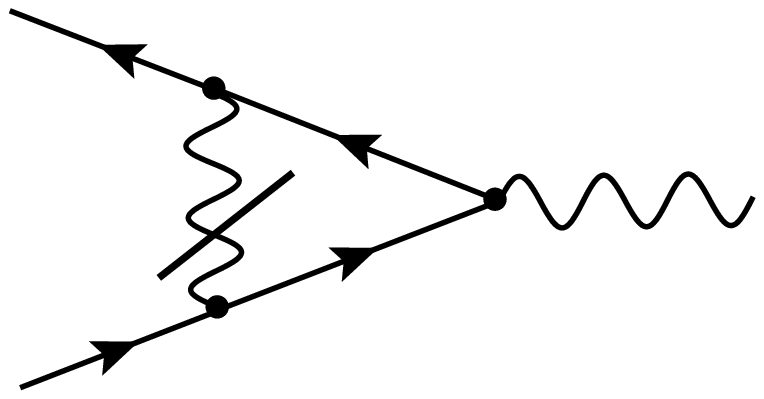}
\caption{The diagrams governing the fRG flow of (a) the fermionic self-energy $\Sigma_f^\Lambda$, (b) the bosonic self-energy $\Sigma_b^\Lambda$, and (c) the Yukawa coupling, $g^\Lambda$. The dash indicates a derivative with respect to the bosonic regulator.}
\label{diagrams}
\end{figure}

The flow of the fermion self energy, corresponding to the diagram shown in Fig.~\ref{diagrams}(a), is given by
\begin{equation}
\partial_{\Lambda} \Sigma_f^{\Lambda}(\omega,{\bf k}) = 3 (g^{\Lambda})^2 \int^{R}_{\Omega{\bf q}} G^{\Lambda}_f (\omega+\Omega,{\bf k}+{\bf q})  D^{\Lambda}_b (\Omega,{\bf q}) \,, \label{fermiselfflow}
\end{equation}
where $\int^{R}_{\Omega{\bf q}} = \int_{\Omega{\bf q}} (-\partial_{\Lambda} R^{\Lambda})\partial_{R^{\Lambda}}$ acts on the bosonic propagator.

The bosonic matrix element of $[ \Gamma_{R}^{(2) \Lambda} + R^{\Lambda}]^{-1}$ is
\begin{equation}
D^{\Lambda}_b (\Omega,{\bf q}) =-  \left({\mathrm{max}(|\textbf{q}|^2,\Lambda^2) + \Sigma_b^{\Lambda}(\Omega,{\bf q})}  \right)^{-1} \,,
\end{equation}
where we have used the regulator $R^{\Lambda} = (\Lambda^2 - |\textbf{q}|^2) \Theta(\Lambda^2 - |\textbf{q}|^2)$, $\Theta$ being the step function. $\Sigma_b^{\Lambda}(\Omega,{\bf q})$ is the bosonic self-energy.  Since the fermionic regulator function has already been set to zero, the fermions na{\"i}vely give no contribution to the flow of the bosonic propagator.  Hence, following Lee, Strack, and Sachdev, we must define $\Sigma_b^{\Lambda} (\Omega,{\bf q})$ via the self-consistency relation
\begin{equation}
\Sigma_b^{\Lambda} (\Omega,{\bf q}) = B^{\Lambda} \frac{|\Omega|}{ |\textbf{q}|} \,, \label{boseself}
\end{equation}
where $B^{\Lambda}$ is the coefficient of the Landau-damping term in the low-frequency expansion, $|\Omega| \ll v_f^{\Lambda} |\textbf{q}| $, of the particle-hole bubble $\Pi_{\rm ph}^{\Lambda} (\Omega,{\bf q})= 2 (g^{\Lambda})^2 \int_{\omega {\bf k}} G^{\Lambda}_f (\omega+\Omega,{\bf k}+{\bf q}) G^{\Lambda}_f (\omega,{\bf k}) $ shown in Fig.~\ref{diagrams}(b).  Since we use a low-frequency expansion, the expression (\ref{boseself}) should contain a step function restricting the self-energy to the low-frequency regime.  However, including it does not materially change our results, so we omit it for simplicity.

The flow of the Yukawa coupling $g^{\Lambda}$ is given by the diagram in Fig.~\ref{diagrams}(c):
\begin{equation}
\partial_{\Lambda} g^{\Lambda} = - (g^{\Lambda} )^3 \int_{\Omega{\bf q}}^R  (G^{\Lambda}_f (\omega+\Omega,{\bf k}+{\bf q}))^2 D_b^{\Lambda}(\Omega,{\bf q}) \Big|_{\tilde{k}=0, \omega =0} \,, \label{yukaflow}
\end{equation}
where we have set external frequencies and momenta to zero ($\tilde{k}=0$ for fermions), i.e.\ we follow only the forward-scattering part of the vertex. The flow equations (\ref{fermiselfflow}) and (\ref{yukaflow}) and the self-consistency relation (\ref{boseself}) completely determine the flow of our truncation of $\Gamma^{\Lambda}_R$.
 
\textit{Parameterizing the flow.} During the flow, the dependences of the fermionic propagator on the frequency and the momentum (as described by $A_{\omega}^\Lambda$ and $A_k^\Lambda$) and the Yukawa coupling ($g^\Lambda$) will all be renormalized.  The dependence of these parameters on $\Lambda$ defines a set of anomalous dimensions:\ $\eta_{\omega}^{\Lambda}$, $\eta_k^{\Lambda}$, and $\eta_{g}^{\Lambda}$.  These are given by
\be
\eta_{\omega}^{\Lambda} = - \frac{\mathrm{d} \ln A_{\omega}^\Lambda}{\mathrm{d} \ln\Lambda}; \,\,\,\,
\eta_k^{\Lambda} = - \frac{\mathrm{d} \ln { A}_k^\Lambda}{\mathrm{d} \ln\Lambda}; \,\,\,\,
\eta_{g}^{\Lambda} = - \frac{\mathrm{d} \ln {g}^\Lambda}{\mathrm{d} \ln\Lambda}. \label{alletas}
\ee
Furthermore we introduce the dimensionless variables
\be
{\tilde B}^\Lambda = \frac{B^\Lambda}{\Lambda^2}; \qquad
{\tilde g}^\Lambda = \frac{g^\Lambda}{\sqrt{\Lambda A_{\omega}^\Lambda A_k^\Lambda}}. \label{alltildes}
\ee

The anomalous dimension $\eta_{\omega}^{\Lambda}$ is determined by the scale-dependence of the correction to the $i\omega$ term in the fermionic self-energy (\ref{fermiself}).  This may be obtained by setting ${\tilde k}$ to zero on the right-hand side of (\ref{fermiselfflow}), taking an $i\omega$-derivative and the limit $\omega \to 0$:
\be
\eta_{\omega}^{\Lambda} = 6 \left( {\tilde g}^\Lambda \right)^2 \! \int_{\bf q}^{\vert {\bf q} \vert < 1} \!\! \int_\Omega \frac{1}{(i\Omega - \vert {\bf q} \vert \cos\theta)^2} 
\frac{1}{\left( v_f^\Lambda {\tilde B}^\Lambda \frac{\vert \Omega \vert}{\vert {\bf q} \vert} + 1 \right)^2} . \label{etafz2}
\ee
The integral in (\ref{etafz2}) depends on the scale only through $v_f^\Lambda {\tilde B}^\Lambda$, so (\ref{etafz2}) defines a function $f(x)$ by $\eta_{\omega}^{\Lambda} =  ({\tilde g}^{\Lambda})^2 f(v_f^\Lambda {\tilde B}^\Lambda)$. $f(x)$ is monotonically increasing with the limiting values $f(x \to 0)=0$ and $f(x \to \infty) = 3/ \pi^2$.

The anomalous dimension $\eta_k^{\Lambda}$ is determined by the correction to the ${\tilde k}$ term in the fermionic self-energy (\ref{fermiself}).  This may be obtained by setting $\omega$ to zero on the right-hand side of (\ref{fermiselfflow}) and then taking a ${\tilde k}$-derivative:
\bea
\eta_k^{\Lambda} & = & 6 \left( {\tilde g}^\Lambda \right)^2 \lim_{\tilde{k} \to 0} \frac{\partial}{\partial {\tilde k}} \Bigg[ \int_{\bf q}^{\vert {\bf q} \vert < 1} \int_\Omega \frac{1}{(i\Omega - [ {\tilde k} + \vert {\bf q} \vert \cos\theta])} \nonumber \\
& & \qquad \qquad \qquad \qquad \left. \times \, \frac{1}{\left( v_f^\Lambda {\tilde B}^\Lambda \frac{\vert \Omega \vert}{\vert {\bf q} \vert} + 1 \right)^2} \Bigg]  \right. . \label{etafa2}
\eea
Again, the part in square brackets depends on the scale only via $v_f^\Lambda {\tilde B}^\Lambda$. The $\tilde{k}$ derivative and the $\Omega$ integration do not commute; this is a consequence of setting the fermionic regulator function to zero. In terms of the function defined by (\ref{etafz2}), $\eta_k^{\Lambda} =  ({\tilde g}^{\Lambda})^2 [ f(v_f^\Lambda {\tilde B}^\Lambda) - f(\infty) ]$.

The particle-hole bubble, Fig.~\ref{diagrams}(b), determines the bosonic renormalization parameter ${\tilde B}^\Lambda$:
\be
{\tilde B}^\Lambda = \frac{k_F}{\pi\Lambda} \frac{\left( {\tilde g}^\Lambda \right)^2}{v_f^\Lambda}. \label{boseselfcons2}
\ee
Finally, we obtain the anomalous dimension of the Yukawa coupling, $\eta_{g}^{\Lambda}$, which is precisely equal to minus one third of the fermionic anomalous dimension at all scales, $\eta_{g}^{\Lambda} = - \eta^\Lambda_\omega / 3$.  This is in sharp contrast to a patch-based treatment, in which $\eta_g^\Lambda$ would be identically zero \cite{mross}.

{\it Low-energy fixed point.} From (\ref{alltildes}), we derive the flow equation:
\be
-\Lambda \partial_{\Lambda } {\tilde g}^\Lambda =  \left( \frac{1}{2} + \eta_{g}^{\Lambda} - \frac{1}{2}\eta_{\omega}^{\Lambda} - \frac{1}{2}\eta_k^{\Lambda} \right) {\tilde g}^\Lambda . \label{flowg}
\ee
At the start of the flow, $A^{\Lambda_{\rm UV}}_\omega=A^{\Lambda_{\rm UV}}_k=1$.  The value of ${\tilde g}^{\Lambda_{\rm UV}}$ does not matter, provided that it is non-zero; this is because, in the absence of the bosonic mass term, there are no relevant operators at the fixed point.  According to (\ref{boseselfcons2}), a non-zero ${\tilde g}^{\Lambda_{\rm UV}}$ implies a non-zero ${\tilde B}^{\Lambda_{\rm UV}}$; in other words, we need to include a small amount of Landau damping even in the bare action to get the flow started.

For a finite-coupling fixed point, the term in the brackets on the right-hand side of (\ref{flowg}) must be zero, and hence the anomalous dimensions at the fixed point obey $1 = - 2 \eta_{g} + \eta_{\omega}+ \eta_k$. Furthermore, according to (\ref{boseselfcons2}) if ${\tilde g}^\Lambda$ remains finite at the fixed point, $v_f^\Lambda {\tilde B}^\Lambda$ must diverge like $\Lambda^{-1}$.  This means that we need only compute the limiting values of the right-hand sides of (\ref{etafz2}) and (\ref{etafa2}) as $v_f^\Lambda {\tilde B}^\Lambda \to \infty$.  Consequently, at the fixed point,
$\eta_{\omega} = - 3 \eta_{g} = 3 ( {\tilde g}/\pi )^2$, while $\eta_k = 0$, which implies that
\be
\eta_{\omega} = - 3 \eta_{g} = \frac{3}{5}; \quad \eta_k= 0; \quad {\tilde g} = \frac{\pi}{\sqrt{5}}. \label{fpvalues}
\ee

The scaling of the bosonic propagator follows from (\ref{boseselfcons2}) and (\ref{fpvalues}), whence ${B}^\Lambda ={\tilde B}^\Lambda \Lambda^2 \sim \Lambda^{1-\eta_\omega+\eta_k}$.  Because we are working in the low-frequency limit ($|\Omega| \ll v_f^{\Lambda} |\textbf{q}| $) we interpret $\Lambda$ as a momentum scale, so that $B^\Lambda \sim \vert {\bf q} \vert^{1-\eta_\omega+\eta_k}$.  This yields a Landau-damping term of the form $\vert \Omega \vert / \vert {\bf q} \vert^{\eta_\omega-\eta_k}$, whence it follows that $z_b=13/5$:
\be
\left( D_b(\Omega,{\bf q}) \right)^{-1} \sim \frac{\vert \Omega \vert}{\vert {\bf q} \vert^{3/5}} + {\bf q}^2. \label{boseprop}
\ee
This corresponds to the surprising result that the magnetic fluctuations at this fixed point have a Landau-damping rate $\Gamma_{\bf q} \sim \vert {\bf q} \vert^{3/5}$, as opposed to the usually expected $\Gamma_{\bf q} \sim \vert {\bf q} \vert$.

The lack of a fermionic regulator in our approach results in an apparent ambiguity when one tries to determine the scaling form of the fermionic Green's function.  One option would be to choose a scaling in which the fermionic and bosonic momenta scale together, as in \cite{drukier}; but in this calculation that appears to be a poor choice for at least two reasons.  First, it yields results that do not agree with the one-loop evaluation of the fermionic self-energy \cite{reizer} when vertex corrections are neglected.  Second, it results in the bosonic and fermionic frequencies' scaling differently as the fixed point is approached.  This suggests that we should rather choose a scaling in which $\omega \sim \Omega \sim \Lambda^{z_b}$; consistency then requires different scaling for the fermionic and bosonic momenta:\ ${\tilde k} \sim \Lambda^2$ whereas $\vert {\bf q} \vert \sim \Lambda$.  This scaling recovers the one-loop self-energy $\Sigma_f \sim \omega^{2/3}$ when vertex corrections are neglected, and furthermore is precisely the scaling that is used in patch theories \cite{polchinski,lee_gauge,met_i,mross,dalidovich,sur}.
The resulting fermionic propagator has the following ${\tilde k}=0$ form at the interacting fixed point:
\be
\left( G_f(\omega,\textbf{k}_F) \right)^{-1} \sim \omega^{1-(\eta_{\omega}/z_b)} \sim \omega^{10/13}, \label{fermiprop1}
\ee
and the following form at zero frequency:
\be
\left( G_f(0,{\textbf k}) \right)^{-1} \sim {\tilde k}^{1-(\eta_k/2)} \sim {\tilde k}. \label{fermiprop2}
\ee
Unlike in ref.~\cite{met_i}, the ${\tilde k}$-dependence of this Green's function is unrenormalized.  However, this is due to the absence in our truncation of certain diagrams \cite{lee_gauge}, included in \cite{met_i,mross,drukier}, that would provoke such a renormalization.

(\ref{fermiprop1}) and (\ref{fermiprop2}) lead to a fermionic dynamical exponent $z_f = z_b/2 = 13/10$. The scaling may also be characterised by a vanishing quasiparticle weight $Z_f^{\Lambda} \sim \Lambda^{\eta_\omega} \sim \Lambda^{3/5}$ and a vanishing Fermi velocity $v_f^{\Lambda} \sim \Lambda^{\eta_\omega-\eta_k} \sim \Lambda^{3/5}$ as $\Lambda \to 0$. (\ref{fermiprop1}) and (\ref{fermiprop2}) demonstrate the non-Fermi-liquid character of the fermions at the fixed point.  A thermodynamic consequence of this is that, in the non-Fermi-liquid regime above the quantum critical point, the specific heat capacity should depend on temperature as $C(T) \sim T^{1/z_f} \sim T^{10/13}$ \cite{sent_rev}.

In summary, the fermionic and bosonic dynamical exponents are given respectively by
\be
z_b = \frac{13}{5}; \qquad
z_f = \frac{z_b}{2} = \frac{13}{10}. \label{dynexp}
\ee
Equations (\ref{boseprop}--\ref{dynexp}) are the main results of this Letter.

{\it Discussion.}  To compare our results with the literature on this problem, we must extend them to the case in which the fermions come in a large number of flavors, $N_f$.  This extension is straightforward:\ $N_f$ appears only in those diagrams containing a fermion loop, so the only place in which it enters is in the determination of $v_f^\Lambda {\tilde B}^\Lambda$.  Since this is in any case infinite at the fixed point, altering the value of $N_f$ makes no difference to our results.

As mentioned above, the $z_f=z_b/2$ scaling that emerges at the fixed point matches that used in patch theories.  However, to be forced to this scaling at the fixed point is not the same as assuming it from the beginning.  We began with a full Fermi surface, not a set of patches, and we find low-energy behavior that is inconsistent with the results of a patch approach.  In particular, the Landau damping of our bosonic propagator is unambiguously renormalized for any $N_f$, unlike in a patch theory where one finds no corrections to the conventional $\vert \Omega \vert / \vert {\bf q} \vert$ form.  This is due to the importance, in our calculation, of the vertex correction (Fig.~\ref{diagrams}(c)):\ $\eta_g = - \eta_\omega/3$ at all stages in the flow, including at the fixed point.

How robust are the numerical values of our exponents to extensions of the truncation scheme used for $\Gamma^\Lambda_R$?  The most obvious term to include would be the bosonic mass:\ in our calculation this was set to zero throughout, but it should more properly be tracked during the flow.  To include the possibility of superconductivity in the vicinity of the quantum critical point, analysing the effect of four-fermion interactions is also important.  Lastly, there is the question of whether the multi-boson interaction vertices, neglected in this treatment, remain well behaved at the fixed point.  Even if they do, we would expect them to alter the anomalous dimensions, and in particular to give a non-zero $\eta_k$ \cite{met_i,mross,drukier}.

Regardless of changes to the exponents that might result from an improved truncation scheme, a clear experimental prediction of our work is that the Landau-damping rate $\Gamma_{\bf q} \sim \vert {\bf q} \vert^{\alpha}$ with $\alpha<1$.  This should be contrasted with the usually expected $\Gamma_{\bf q} \sim \vert {\bf q} \vert$, which is in reality only perturbatively valid.  This calls to mind the measurements taken some years ago on UGe$_2$ \cite{huxley} and more recently on other similar compounds \cite{stock}, which when fitted with a $\vert {\bf q} \vert$ form show a damping rate that does not extrapolate to zero in the $\vert {\bf q} \vert \to 0$ limit.  At least two theories have been proposed \cite{mineev,betouras} in which it would not be expected to.  However, our work raises an interesting alternative possibility:\ that the data should be fitted with a free exponent.

UGe$_2$, however, is not a cleanly 2D material, which complicates the application of our theory.  The properties of the quasi-2D compound YFe$_2$Al$_{10}$ have recently been measured \cite{aronson}, and may provide a closer fit.  In particular, it is reported that the Sommerfeld coefficient $C/T \sim \ln T$ near the quantum critical point.  Our specific fixed-point theory predicts $C/T \sim T^{-3/13}$; more generally, a $T^{-\beta}$ behavior with $\beta$ small compared to 1 appears to be an equally good fit to the data over the temperature range where scaling behavior was observed.  Single-crystal neutron scattering data on YFe$_2$Al$_{10}$ are not yet available, but if this material is indeed a 2D nearly ferromagnetic metal --- as the authors of \cite{aronson} claim --- then we would predict $\Gamma_{\bf q} \sim \vert {\bf q} \vert^{\alpha}$ with $\alpha < 1$.

Finally, we discuss some of the implications of our work for the antiferromagnetic critical point. Our analysis above sheds some light on an unresolved issue in the paper by Lee, Strack, and Sachdev, viz.\ why $z_b=z_f$.  The fact that the scaling of the bosonic self-energy is fully determined by the low-frequency expansion of the particle-hole bubble enforces $z_b=z_f$ at the antiferromagnetic fixed point.  This is a trivial variation of our argument above that $z_b=2z_f$ in the ferromagnetic case.  In the antiferromagnetic case, the fixed point presented in \cite{af_frg} is for $N_f=1$ while the diagrammatics are valid for $N_f \gg 1$, so we cannot directly assess whether they agree or not.  The extension of the antiferromagnetic fRG analysis to the $N_f \gg 1$ case will result in changes to the anomalous dimensions at the fixed point; this would provide a useful benchmark of fRG against other methods.

{\it Acknowledgments.} We thank A.V.\ Chubukov for useful discussions, A.M.\ Tsvelik for making us aware of the experimental work on YFe$_2$Al$_{10}$, M.C.\ Aronson for showing us ref.\ \cite{aronson} prior to its publication, and A.D.\ Huxley for drawing the case of UGe$_2$ to our attention.  Financial support from the EPSRC (UK) under grants EP/G03673X/1 (SPR) and EP/I031014/1 (CAH) is acknowledged.


\bibliography{prl_frg}


\end{document}